%% file: cuore0_performance.tex
\documentclass[twocolumn,epjc3]{svjour3} % twocolumn
\RequirePackage{graphicx,subfig,multirow,ulem}
\RequirePackage{mathptmx} % use Times fonts if available on your TeX system
\RequirePackage[numbers,sort&compress]{natbib}
\RequirePackage[colorlinks,citecolor=blue,urlcolor=blue,linkcolor=blue]{hyperref}

\newcommand{\1}[1]{~ \mathrm{#1}} % unit(y ;-)
\newcommand{\n}[1]{\mathrm{#1}} % normal (roman) text in math mode

\newcommand{\optprefix}[1]{#1\mbox{-}\ignorespaces}
\journalname{Eur. Phys. J. C}

\begin{document}

\title{Initial performance of the \optprefix{CUORE}0 experiment}

\input{author-20140202.tex}

\maketitle

\begin{abstract}
\optprefix{CUORE}0 is a cryogenic detector that uses an array of
tellurium dioxide bolometers to search for neutrinoless double-beta
decay of $\n{^{130}Te}$. We present the first data analysis with $7.1
\1{kg \cdot y}$ of total $\n{TeO_2}$ exposure focusing on background
measurements and energy resolution. The background rates in the
neutrinoless double-beta decay region of interest (2.47 to $2.57
\1{MeV}$) and in the $\alpha$ background-dominated region (2.70 to
$3.90 \1{MeV}$) have been measured to be ${0.071\pm 0.011}$ and
${0.019\pm 0.002} \1{counts/(keV\cdot kg\cdot y)}$, respectively. The
latter result represents a factor of 6 improvement from a predecessor
experiment, Cuoricino. The results verify our understanding of the
background sources in \optprefix{CUORE}0, which is the basis of
extrapolations to the full CUORE detector.  The obtained energy
resolution (full width at half maximum) in the region of interest is
$5.7 \1{keV}$. Based on the measured background rate and energy
resolution in the region of interest, \optprefix{CUORE}0 half-life
sensitivity is expected to surpass the observed lower bound of
Cuoricino with one year of live time.
\end{abstract}



%%%%%%%%% Section I: Introduction %%%%%%%%%
\section{Introduction}
\label{sec:introduction}

Neutrinoless double-beta decay ($0\nu\n{DBD}$) is a hypothetical
lepton number violating process in which two neutrons in an atomic
nucleus simultaneously decay to two protons, two electrons, and no
electron-antineutrinos: ${(A,Z)\rightarrow(A,Z+2) +
  2e^-}$. Observation of $0\nu\n{DBD}$ would establish the Majorana
nature of the neutrino, i.e., that the neutrino is its own
antiparticle, and may provide insights on the neutrino mass scale and
mass hierarchy, depending on $0\nu\n{DBD}$ rate or rate limit
(cf.~\cite{Avignone:2007fu}). The experimental signature for
$0\nu\n{DBD}$ is a peak at the $0\nu\n{DBD}$ \optprefix{Q}value in the
two-electron energy sum spectrum. Several recent experiments have
reported new limits on the $0\nu\n{DBD}$ half-life of
$\n{^{136}Xe}$~\cite{Auger:2012ar, Gando:2012zm} and
$\n{^{76}Ge}$~\cite{Agostini:2013mzu}. For comparison between
experiments, half-life limits of different isotopes are usually
converted to limits on the effective Majorana mass. This conversion,
however, takes into account the phase space factors and nuclear matrix
elements, the latter of which introduce large uncertainties from
different model calculations~\cite{Menendez:2008jp,
  PhysRevC.87.014301, Rodriguez:2010mn, Fang:2011da, Faessler:2012ku,
  suhonen_review_2012, Barea:2013bz}. The current $0\nu\n{DBD}$
half-life limit for $\n{^{130}Te}$ was set by Cuoricino at
$2.8\times10^{24}\1{y}$ (90\% C.L.)~\cite{Andreotti:2010vj}.

\optprefix{CUORE}0 is a cryogenic detector that uses an array of
$\n{TeO_2}$ bolometers to search for $0\nu\n{DBD}$ in the
$\n{^{130}Te}$ of the bolometers themselves. $\n{^{130}Te}$ is an
attractive isotope for a $0\nu\n{DBD}$ search because of its
relatively high \optprefix{Q}value at $2528 \1{keV}$~\cite{Redshaw:2009cf,
  Scielzo:2009nh,Rahaman2011412} and its very high natural isotopic
abundance at 34.2\%~\cite{Fehr:2004aa}. Cryogenic bolometers measure
energy through a rise in the temperature of the detector and have
energy resolutions comparable to high purity Ge detectors: for
\optprefix{CUORE}0 style bolometers, the energy resolution (full width
at half maximum, FWHM) is typically 0.2\% at the $0\nu\n{DBD}$
\optprefix{Q}value. $0\nu\n{DBD}$ data taking with \optprefix{CUORE}0 began in March
2013.

\optprefix{CUORE}0 also serves as a technical prototype for CUORE
(Cryogenic Underground Observatory of Rare
Events)~\cite{Ardito:2005ar}, which will consist of 19 towers
identical to the single CUORE-0 tower. \optprefix{CUORE}0 is the first
tower produced on the CUORE assembly line, and its successful
commissioning represents a major milestone towards CUORE. CUORE is in
the advanced stages of detector construction at the time of this
writing and is scheduled to begin data taking in 2015.

%%%% Section II: CUORE-0 Detector %%%%
\section{\optprefix{CUORE}0 detector}
\label{sec:cuore0_det}

\begin{figure}[b]
%\begin{center}
 \begin{tabular}{l c}
 \subfloat[]{\label{fig:tower}\includegraphics[width=0.155\columnwidth]{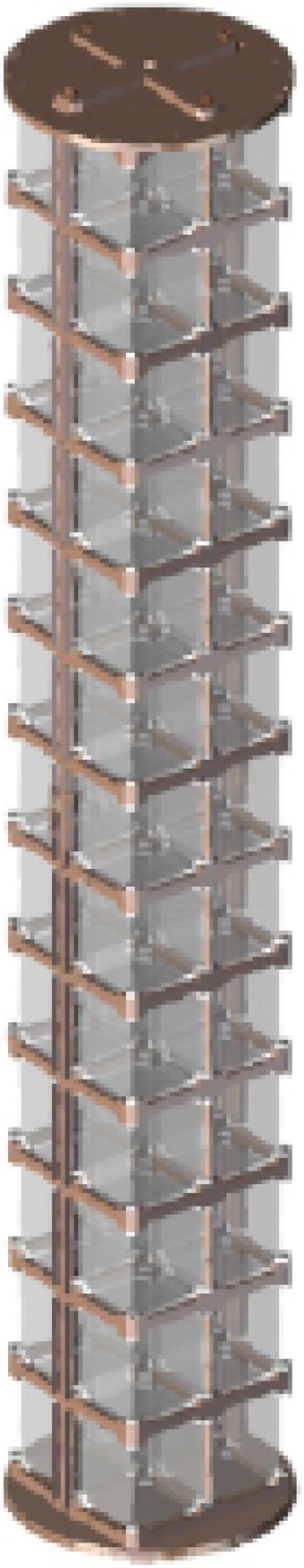}} &
 \multirow{-27}[6]{*}{\subfloat[]{\label{fig:illustration}\includegraphics[width=0.43\columnwidth]{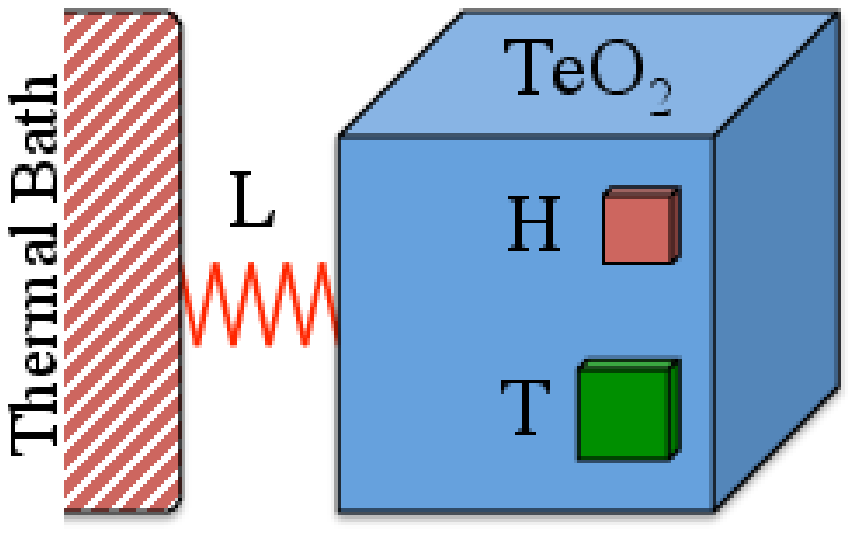}}}\\
 & \multirow{-11}[6]{*}{\subfloat[]{\label{fig:bol_sig}\includegraphics[width=0.76\columnwidth]{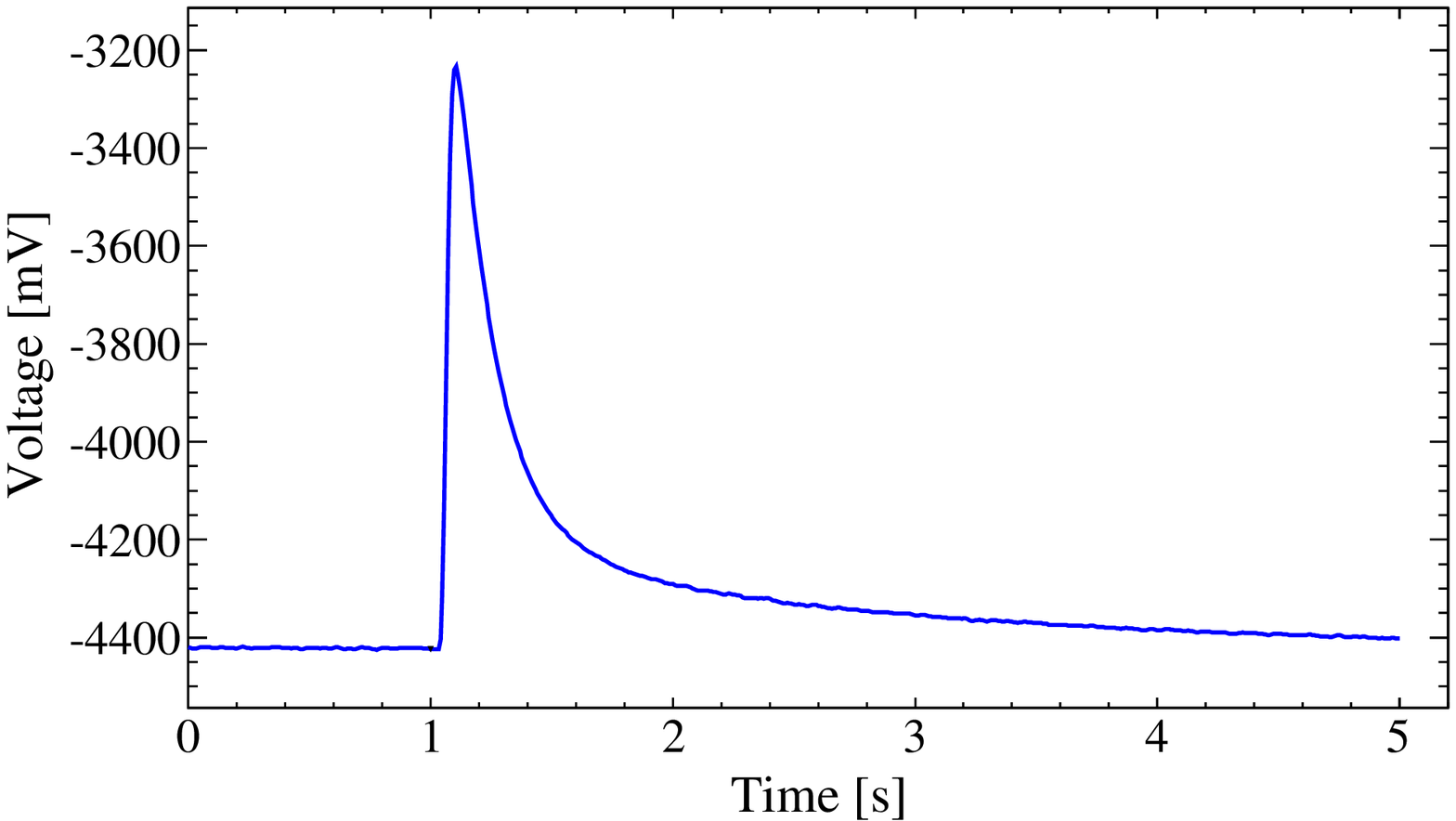}}} \\
 \end{tabular}
%\end{center}
\caption{(a) \optprefix{CUORE}0 tower array rendering. The tower
  consists of 13 planes of 4 crystals, mounted in the frame made of
  copper. (b) Schematic of a single \optprefix{CUORE}0 bolometer
  showing the thermistor (T), the heater (H), and the weak
  thermal link (L) between $\n{TeO_2}$ crystal and copper thermal bath
  (not to scale).  (c) An example of a bolometer signal with the
  energy of approximately $2615\1{keV}$. The rise and fall times of
  this signal are $0.04\1{s}$ (defined as the time for the pulse
  amplitude to evolve from 10\% to 90\% of its maximum) and
  $0.26\1{s}$ (90\% to 30\% of maximum), respectively.}
\end{figure}

\optprefix{CUORE}0 is a single tower of 52 $\n{TeO_2}$ crystal
bolometers operating at a typical base temperature of $13 - 15 \1{mK}$
in Hall A of the Laboratori Nazionali del Gran Sasso (LNGS)
underground facility in Italy. As shown in Figure~\ref{fig:tower}, the
tower consists of 13 planes of four $5\times5\times5\1{cm^3}$
crystals, held securely inside a copper frame by specially designed
polytetrafluoroethylene (PTFE) brackets.  The copper frame serves as a
thermal bath to cool the crystals through the weak thermal coupling
provided by PTFE. Each crystal weighs $750\1{g}$, which results in a
total detector mass of $39\1{kg}$ and a total $^{130}\n{Te}$ mass of
$11\1{kg}$.  Each crystal is instrumented with a single neutron
transmutation doped (NTD) germanium thermistor for the signal readout
(see Figure~\ref{fig:illustration}). The typical signal amplitude
$\Delta T/\Delta E$ is $10 - 20\1{\mu
  K}$/MeV. Figure~\ref{fig:bol_sig} shows an example of a bolometer
signal. Additionally, one silicon Joule
heater~\cite{Alessandrello:1998bf} is also glued to the crystal for
the offline correction of thermal gain drift caused by temperature
variation of the individual bolometer.

We put significant effort into the selection and handling of the
detector materials with the objective of minimizing the background
contamination for \optprefix{CUORE}0. In collaboration with the
$\n{TeO_2}$ crystal grower at the Shanghai Institute of Ceramics,
Chinese Academy of Sciences, we developed a radiopurity control
protocol~\cite{Arnaboldi:2010fj} to limit bulk and surface
contaminations introduced in crystal production. Only materials
certified for radiopurity were used to grow the crystals. After
production, the crystals were transported to LNGS at sea level to
minimize cosmogenic activation. Upon arrival at LNGS, a few crystals
from each batch were instrumented as bolometers for characterization
tests.  For $\n{^{238}U}$ ($\n{^{232}Th}$) decay chain, the measured
bulk and surface contaminations are less than
$6.7\times10^{-7}\1{Bq/kg}$ ($8.4\times10^{-7}\1{Bq/kg}$) and
$8.9\times10^{-9}\1{Bq/cm^2}$ ($2.0\times10^{-9}\1{Bq/cm^2}$) at
$90\1{\%}$ C.L., respectively~\cite{Alessandria:2011vj}.  Material
screening of small parts, including NTD thermistors and silicon
heaters, indicates that their radioactive content contributes to less
than 10\% of the total background in the $0\nu\n{DBD}$ region of
interest (ROI).

Based on the experience of Cuoricino, we expect the most significant
background contributions to come from the tower frame and the
surrounding cylindrical thermal shield, both of which are made from
radiopure electrolytic tough pitch copper~\cite{Aurubis}. Relative to
Cuoricino, the total mass and surface area of the tower frame of
\optprefix{CUORE}0 was reduced by a factor of 2.3 and 1.8,
respectively. Monte Carlo studies predict a factor of 1.3 decrease in
$\alpha$ background from the thermal shield arriving at the crystals
due to the change in the geometry~\cite{Artusa:2014aa}.  To further
mitigate the surface contamination of the copper structure, we tested
three surface treatment techniques~\cite{Alessandria:2012zp} and chose
a series of tumbling, electropolishing, chemical etching, and
magnetron plasma etching for the surface treatment. The upper limit of
the surface contamination of the cleaned copper was measured in R\&D
bolometers to be $1.3\times 10^{-7}\1{Bq/cm^2}$ (90\% C.L.) for both
$\n{^{238}U}$ and $\n{^{232}Th}$~\cite{Alessandria:2012zp}.

The \optprefix{CUORE}0 detector assembly procedure was designed to
minimize the recontamination of clean components. Tower assembly takes
place in a dedicated class 1000 clean room in Hall A of the LNGS
underground facility. To minimize exposure to radon (and radon
progeny) in air, all steps of the assembly were performed under
nitrogen atmosphere inside glove boxes~\cite{Clemenza_radon_2011}. All
tools used inside the glove boxes, and especially those that would
physically touch the detector components, were cleaned and certified
for radiopurity. The assembled tower was enclosed in a copper thermal
shield and mounted in the Cuoricino cryostat. To minimize exposure to
the environment during mounting to the cryostat, mounting was
performed in the Cuoricino clean room, and the tower was kept under
nitrogen flux for as long as possible.

\optprefix{CUORE}0 uses for the first time flexible printed circuit
board (PCB) cables and \textit{in situ} wire bonding for electrical
wiring of the tower. This is one of the major upgrades that
significantly improved the robustness of bolometer readout wiring
compared to the Cuoricino design.  A set of flexible PCB cables with
copper traces~\cite{Brofferio:2013cya, Andreotti:2009zza} was attached
to the copper frame from the bottom plane to the top.  The lower ends
of the PCB copper traces were bonded to the metal contact pads of the
thermistors and heaters using $25\1{\mu m}$ diameter gold wires. The
upper ends of the PCB cables were connected through another
custom-made flexible PCB at the $10\1{mK}$ plate to a set of Manganin
twisted pair flat ribbon cables running un-interrupted to the
feedthroughs on the top plate of the cryostat. Overall, only
3 bolometers (6\%) are not fully functional from the loss of 1
thermistor and 2 heaters. The two heater-less bolometers can be used
in non-standard analysis without thermal gain correction in the
future.

\optprefix{CUORE}0 is operated in the same cryostat, uses the same
external lead and borated-polyethylene neutron shielding, and is
enclosed in the same Faraday cage that was used for
Cuoricino~\cite{Andreotti:2010vj,Arnaboldi:2008ds}.  The front-end
electronics~\cite{Arnaboldi:2010zz,Arnaboldi:2004jj,Arnaboldi:2002aa}
and data acquisition hardware are also identical to those used in
Cuoricino. We implemented a new automated bias voltage scanning
algorithm to locate the optimal working point that maximizes the
signal-to-noise ratio (SNR). The bolometer signals are amplified and
then filtered with six-pole Bessel low-pass filters. Subsequently,
signals are digitized by two 32-channel National Instruments PXI
analog-to-digital converters with a $125\1{S/s}$ sampling rate, 18-bit
resolution, and $21\1{V}$ full scale. All samples are stored
continuously on disk. Afterwards, in almost real-time, a 
constant fraction analysis trigger identifies triggered pulses with
626 sampling points ($5.008\1{s}$), including a pre-trigger segment of
125 samples. Each bolometer has an independent trigger threshold,
ranging from $50$ to $100 \1{keV}$. In addition to the signal
triggers, each bolometer is pulsed periodically at $300\1{s}$
intervals with a fixed and known energy through the heater. These
``pulser'' events are used to monitor and correct the gain of the
bolometers~\cite{Arnaboldi:2003yp}. Finally, a baseline trigger
identifies a baseline pulse every $200\1{s}$ to provide snapshots of the
detector working temperatures and noise spectra.

\optprefix{CUORE}0 data are grouped into ``data sets''. Each data set
consists of a set of initial calibration runs, a series of physics
runs, and a set of final calibration runs.  Calibration data refers to
the sum of all calibration runs, while background data refers to the
sum of all physics runs to search for $0\nu\n{DBD}$.  During
calibrations, the detector is irradiated using two thoriated tungsten
wires, each with a $\n{^{232}Th}$ activity of $50 \1{Bq}$. The wires
are inserted into two vertical tubes on opposite sides of the tower
that run between the outer vacuum chamber and the external lead
shielding. We calibrate each channel using $\gamma$ rays from daughter
nuclei of $\n{^{232}Th}$ in the energy range from $511$ to
$2615\1{keV}$. The signal rates on each bolometer for the calibration
and background data are $60 - 70$ and $0.5 - 1.0\1{mHz}$,
respectively.

\section{\optprefix{CUORE}0 performance and background}
\label{sec:cuore0_performance}
The \optprefix{CUORE}0 data reported in this article was collected
between March and September 2013, with interruptions for dilution
refrigerator maintenance.  To account for temporary degraded
performances on each individual bolometer due to large baseline
excursions or elevated noise levels, we reject low-quality data
intervals on a channel-by-channel basis.  Consequently, the total
exposure is obtained by summing the individual exposures of each
bolometer. The accumulated $\n{TeO_2}$ exposure on 49 fully active
channels is $7.1 \1{kg\cdot y}$ for a $^{130}\n{Te}$ isotopic exposure
of $2.0 \1{kg \cdot y}$, excluding all low-quality data intervals.

\optprefix{CUORE}0 data analysis follows the same procedure as was
used for Cuoricino~\cite{Andreotti:2010vj}. This includes amplitude
evaluation, gain correction, energy calibration, and time coincidence
analysis among the bolometers. Pulse amplitude is evaluated by first
maximizing the SNR with an optimum filter. Fourier components of each
pulse are weighted at each frequency by the expected SNR, which is
calculated for each channel with an average pulse of $2615 \1{keV}$
$\gamma$ rays and the average power spectra of the noise events.  For
the gain correction of each bolometer, the amplitudes of pulser events
are fit against their baseline voltages to determine the gain
dependence on temperature, and this temperature dependence is backed
out for each signal pulse. For the energy calibration, we use a third
order polynomial fit in the energy range from 0 to $3.9\1{MeV}$ since
the relationship between energy and stabilized amplitude is found to
be slightly nonlinear. The deviation from a linear fit is less than
$10 \1{keV}$ at the $2615\1{keV}$ peak. If any two or more crystals
register signal pulses within $100\1{ms}$ of each other, the events
are tagged as coincidence events. These are mostly attributed to
backgrounds such as Compton-scattered $\gamma$ rays or $\alpha$ decays
on the surface of two adjacent crystals.

\begin{figure}[tb]
\begin{center}
 \includegraphics[width=1.0\columnwidth]{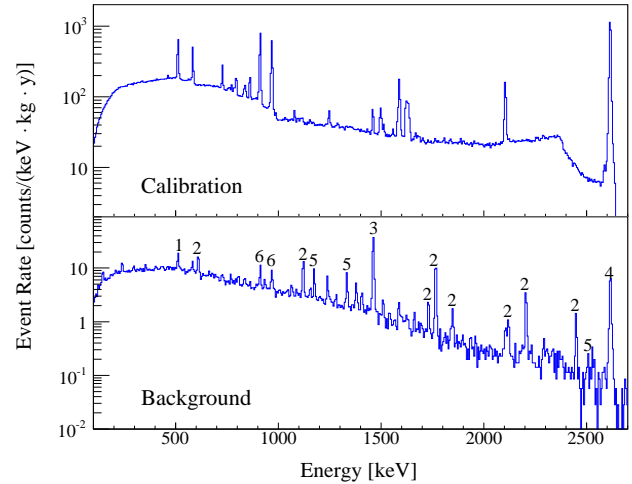}
 \caption{\optprefix{CUORE}0 calibration (top panel) and 
   background spectrum (bottom panel) over the data taking
   period presented in 
   this work. $\gamma$-ray peaks from known radioactive sources in the
   background spectrum are labeled as follows: (1)~$e^+e^-$
   annihilation; (2)~$\n{^{214}Bi}$; (3)~$\n{^{40}K}$;
   (4)~$\n{^{208}Tl}$; (5)~${}^{60}\n{Co}$; and (6)~$\n{^{228}Ac}$. }
\label{fig:gamma_bkg_cal}
\end{center}
\end{figure}

The event selection criteria can be categorized as follows: basic data
quality, pile-up, pulse shape, and anti-coincidence. The basic data
quality cut rejects events within low-quality data intervals, as
mentioned in the beginning of the section.  The pile-up cut requires
that only one pulse exists in a $7.1\1{s}$ window around the measured
trigger time (see Figure~\ref{fig:bol_sig}). Due to the relatively
long rise and decay times of a pulse and negligible pulse shape
dependence on energy at energies above $1\1{MeV}$, the pulse shape of
the possible ${}^{130}\textrm{Te}$ $0\nu\textrm{DBD}$ signal is
expected to be similar to that obtained from the $2615\1{keV}$
$\gamma$-ray peaks. Therefore, the pulse shape cut requires that the
signal shape is comparable to that obtained from the average pulse
recorded with $2615 \1{keV}$ $\gamma$-ray events, and that the
pre-trigger baseline slope is smaller than $0.1\1{mV/Sample}$. The
anti-coincidence cut requires that no other pulse in coincidence is
recorded in the entire tower.

We evaluate the selection efficiency mainly using the $2615\1{keV}$
$\gamma$-ray peak since it offers sufficient statistics at the energy
closest to the ROI. However, since the $2615\1{keV}$ $\gamma$-ray
events occasionally occur in coincidence with other physical events,
the efficiency of anti-coincidence cut was evaluated using the
$1461\1{keV}$ $\gamma$ rays from ${}^{40}\n{K}$ decay, which are truly
individual events. The selection efficiency was averaged over all
active channels. The efficiency was obtained by first evaluating the
slowly varying background rate under the peak by counting the number
of events in the energy regions between 3 and $15\sigma$ above and
below the peak. The background rate was then subtracted from the peak
rate which was measured by counting the number of events within $\pm
3\sigma$ of the peak. The result was cross-checked by fitting the
combined peak and background region ($\pm15\sigma$) with a Gaussian
plus linear function. The difference between the two methods was
integrated as the systematic uncertainty of the selection
efficiency. The obtained efficiency is ${92.9 \pm 1.8\%}$, which is
the efficiency of all cuts other than the anti-coincidence cut,
obtained from the $2615\1{keV}$ $\gamma$-ray peak, multiplied by the
efficiency of the anti-coincidence cut, obtained from the
$1461\1{keV}$ peak after applying all other cuts, as described above.
Since we are considering only single crystal events, we must include
the confinement efficiency, i.e. the probability that both
$0\nu\n{DBD}$ electrons are contained inside the single crystal. This
probability has been estimated by simulation to be ${87.4 \pm
  1.1\%}$~\cite{Andreotti:2010vj}.  Taking into account the ${99.00
  \pm 0.01\%}$ signal trigger efficiency, which is evaluated on
pulser events, the total $0\nu\n{DBD}$ detection efficiency of
\optprefix{CUORE}0 is ${80.4\pm 1.9\%}$. This result is compatible
with the value obtained from Cuoricino, which was found to be
${82.8\pm 1.1\%}$~\cite{Andreotti:2010vj}.

The top panel in Figure~\ref{fig:gamma_bkg_cal} shows the energy
spectrum obtained using the $^{232}\n{Th}$ calibration source. The
spectrum is the sum of all 49 fully functional channels.  The
histogram in the bottom panel shows the background spectrum of
\optprefix{CUORE}0 for the analysis presented. The presence of the
pronounced $\gamma$-ray peaks from $^{214}\n{Bi}$ decay, a daughter
nucleus of $^{222}\n{Rn}$, is attributed to the inclusion of data
taken without the nitrogen gas purge in the Faraday cage around the
dilution refrigerator. The nitrogen purge effectively suppresses
$^{214}\n{Bi}$ $\gamma$-ray intensities by more than a factor of
5. The energy resolution in the ROI is defined as the FWHM of $2615
\1{keV}$ $\gamma$-ray peak, determined by a fit to
the summed background spectrum of all fully functional channels. The median
resolution of the individual channels in the calibration data is
$6.0 \1{keV}$, with a mean of $6.8\1{keV}$ and a root mean square
deviation of $2.1\1{keV}$.  The relatively larger mean and root mean
square deviation are attributed to a few underperforming channels. 
The energy resolution of calibration data slightly deteriorates due to accidental
coincidence events and a higher overall noise level compared to background data.

When compared to Cuoricino, one new noise contribution is correlated
microphonic noise on multiple channels introduced by the new flexible
PCB wiring. The vibration of one PCB cable might introduce common-mode
noise in all the channels on that cable, which is apparent in the
low-frequency part of the signal band and degrades the energy
resolution. However, even with the correlated noise in play, the
energy resolution of $5.7 \1{keV}$ (FWHM) is better, on average, than
that of Cuoricino. Furthermore, we have on-going studies seeking to
improve the energy resolution by regressing the correlated noise out
of the bolometer signals~\cite{ManciniTerracciano:2012fq}.

As of this writing, we have kept the $0\nu\n{DBD}$ region blinded.
Our blinding procedure is a form of \textit{data salting}, where we
randomly exchange a blinded fraction of events within $\pm 10\1{keV}$
of the $2615\1{keV}$ $\gamma$-ray peak with events within $\pm
10\1{keV}$ of the $0\nu\n{DBD}$ \optprefix{Q}value. The exchange probability
varies between 1 and 3\% and is randomized run by run. Since the
number of $2615 \1{keV}$ $\gamma$-ray events is much larger than that
of possible $0\nu\n{DBD}$ events, the blinding algorithm produces an
artificial peak around the $0\nu\n{DBD}$ \optprefix{Q}value and blinds the real
$0\nu\n{DBD}$ rate of ${}^{130}\n{Te}$. This method of blinding the
data preserves the integrity of the possible $0\nu\n{DBD}$ events
while maintaining the spectral characteristics with measured energy
resolution and introducing no discontinuities in the spectrum.

\begin{figure}[!tb]
\begin{center}
 \includegraphics[width=1.0\columnwidth]{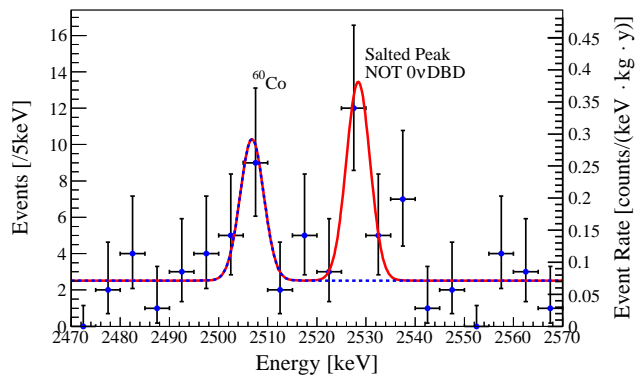}
 \caption{(color online) Blinded energy spectrum and $0\nu\n{DBD}$
   global fit in the region of interest. The unbinned
   maximum-likelihood fit is shown in solid red line. The dotted blue
   line illustrates the peak from ${}^{60}\n{Co}$ and the linear
   background only to highlight the difference between salted
   $0\nu\n{DBD}$ peak and background. The flat background from the fit
   is ${0.071 \pm 0.011
     \1{counts/(keV\cdot kg\cdot y)}}$. $\n{^{60}Co}$ peak position, salted peak position, 
   and rate are ${2506.8 \pm 1.1\1{keV}}$, ${2528.4 \pm 1.0\1{keV}}$, and ${1.3\pm 0.5
     \1{counts/(keV\cdot kg\cdot y)}}$, respectively.}
\label{fig:globalNDBDFit}
\end{center}
\end{figure}

The background rate in the ROI is evaluated using the blinded
spectrum in the energy range ${2470-2570\1{keV}}$. This region
includes the $\n{^{60}Co}$ sum-peak at $2506 \1{keV}$ and the salted
peak at the $0\nu\n{DBD}$ \optprefix{Q}value, as shown in
Figure~\ref{fig:globalNDBDFit}. We use an unbinned maximum-likelihood
fit to estimate the background rate in the ROI. The likelihood
function consists of the sum of a $\n{^{60}Co}$ Gaussian peak, a
salted $0\nu\n{DBD}$ Gaussian peak, and a flat background. In the fit,
the mean of the $\n{^{60}Co}$ peak is initialized to $2506 \1{keV}$
and the mean of the salted $0\nu\n{DBD}$ peak at $2528 \1{keV}$.  The FWHM of
both peaks is fixed to the detector resolution at $5.7 \1{keV}$. As
shown in Figure~\ref{fig:globalNDBDFit}, the fit reveals that the
overall background rate in the ROI is ${0.071 \pm 0.011 ~\n{(stat)}
  ~\1{counts/(keV\cdot kg\cdot y)}}$. For comparison, the background rate of the
Cuoricino crystals with the same dimension is ${0.153 \pm 0.006
  \1{counts/(keV\cdot kg\cdot y)}}$. Systematic uncertainties arising from
background shape are studied by comparing a constant and a linear
background models, and are found to be less than 3\%. The systematic
contribution from the uncertainty in energy calibration is less than
1\%.

\begin{figure}[tb]
\begin{center}
 \includegraphics[width=1.0\columnwidth]{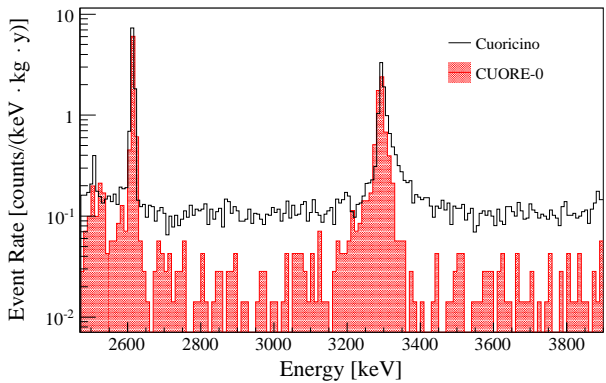}
 \caption{(color online) Background spectrum of \optprefix{CUORE}0
   (red with shades) and Cuoricino (black) in the region dominated by
   degraded $\alpha$ particles. The figure shows reduction of the flat
   background caused by degraded $\alpha$ particles in the energy
   region of [2.7 - 3.1] and $[3.4 - 3.9] \1{MeV}$.}
\label{fig:alpha_bg_comp_cuore0_cuoricino}
\end{center}
\end{figure}

The two major sources of background in the ROI are degraded $\alpha$
particles from surface contamination on the detector components and
$\gamma$ rays that originate from the cryostat. Degraded
$\alpha$ particles with a decay energy of 4 to $8 \1{MeV}$ may deposit
part of their energy in the $0\nu\n{DBD}$ ROI. These $\alpha$ events
form a continuous energy spectrum extending from their decay energy to
well below $0\nu\n{DBD}$ region. The $\alpha$ background rate in the
ROI is estimated by counting events in the ``$\alpha$ flat continuum
region'', which is defined to be from $2.7$ to $3.9\1{MeV}$ (excluding
the $\n{^{190}Pt}$ peak region from 3.1 to $3.4\1{MeV}$). This energy
range is above almost all naturally occurring $\gamma$ rays, in
particular the $2615 \1{keV}$ $\gamma$ rays from ${}^{208}\n{Tl}$
decay.  Figure~\ref{fig:alpha_bg_comp_cuore0_cuoricino} shows the
background energy spectrum of \optprefix{CUORE}0 (shaded red) and
Cuoricino (black). The measured rate for \optprefix{CUORE}0 is $0.019\pm 0.002
\1{counts/(keV\cdot kg\cdot y)}$, which improves on the Cuoricino result
($0.110\pm 0.001 \1{counts/(keV\cdot kg\cdot y)}$) by a factor of 6.

The $\gamma$-ray background in the ROI is predominantly
Compton-scattered $2615 \1{keV}$ $\gamma$ rays originating from
${}^{208}\n{Tl}$ in the cryostat. Since \optprefix{CUORE}0 is hosted
in the same cryostat as was used for Cuoricino, the $\gamma$-ray
background is expected to be similar. The $\gamma$-ray background is
estimated as the difference between overall background in the ROI and
the degraded $\alpha$ background in the continuum. The measured
$\gamma$-ray backgrounds of \optprefix{CUORE}0 and Cuoricino are
indeed compatible~\cite{Andreotti:2010vj}, consistent with the
hypothesis that the background in the ROI is composed of $\gamma$ rays
from the cryostat and degraded $\alpha$ particles.
 
\section{Projected sensitivity of \optprefix{CUORE}0}
\label{sec:cuore0_sensitivity}

\begin{figure}[tbp]
 \begin{center}
 \includegraphics[width=1.0\columnwidth]{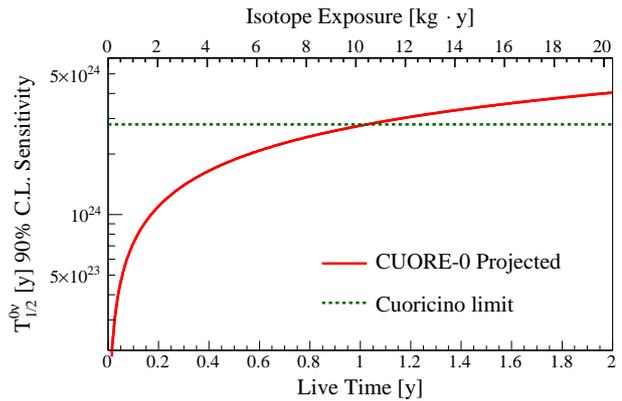}
 \end{center}
 \caption{(color online) Sensitivity of \optprefix{CUORE}0 with the
   measured background rate in the ROI of ${0.071
     \1{counts/(keV\cdot kg\cdot y)}}$ and energy resolution of $5.7 \1{keV}$
   FWHM. The \optprefix{CUORE}0 sensitivity is expected to surpass
   that of Cuoricino with one year of live time.}
 \label{fig:cuore0_sensitivity}
\end{figure}

Using the measured background rate and energy resolution of the
$2615\1{keV}$ $\gamma$-ray peak, we obtain the \optprefix{CUORE}0
sensititvity with the approach outlined
in~\cite{Alessandria:2011rc}. With the excellent energy resolution, we
construct a single-bin counting experiment with a $5.7 \1{keV}$ bin
centered at the $0\nu\n{DBD}$ \optprefix{Q}value. The sensitivity is
obtained by comparing the expected number of signal events with
Poissonian fluctuations from the expected background rate in this
bin. Figure~\ref{fig:cuore0_sensitivity} shows the 90\%
C.L. sensitivity of \optprefix{CUORE}0. With one year of live time, or
$11\1{kg\cdot y}$ isotope exposure, \optprefix{CUORE}0 is expected to
surpass the $^{130}\n{Te}$ $0\nu\n{DBD}$ half-life sensitivity
achieved by Cuoricino, $2.8 \times 10^{24} \1{y}$.

\section{Summary and outlook}
\label{sec:summary}
We present the energy resolution and background measurements of
\optprefix{CUORE}0 detector from from the $7.1 \1{kg \cdot y}$
exposure accumulated up to September 2013. The measured $5.7\1{keV}$
FWHM in the $0\nu\n{DBD}$ ROI represents a slight improvement over
Cuoricino and validates the \optprefix{CUORE}0 wiring scheme and
assembly procedure. The background rates have been measured to be
${0.071 \pm 0.011 \1{counts/(keV\cdot kg\cdot y)}}$ in the ROI and ${0.019 \pm
  0.002 \1{counts/(keV\cdot kg\cdot y)}}$ in the $\alpha$ continuum region. These
results are a factor of 2 and 6 improvement compared to Cuoricino, due
to more rigorous copper surface treatment, improved crystal production
and treatment protocols, and more stringent assembly procedures in the
clean environment. The \optprefix{CUORE}0 sensitivity is expected to
surpass that of Cuoricino with one year of live time.

As a technical prototype for CUORE, \optprefix{CUORE}0 demonstrates
the feasibility of instrumenting an ultra-pure ton-scale bolometer
array with 988 channels.  By enhancing the procedure of the on-going
CUORE assembly, we have improved assembly success rate to close to
100\%, which is a crucial achievement for large arrays such as
CUORE. We have started implementing the noise decorrelation algorithms
into the \optprefix{CUORE}0/CUORE data analysis package, with the aim
of further improving energy resolution. \optprefix{CUORE}0 reconfirms
the effectiveness of the copper cleaning technique and clean assembly
procedure developed for CUORE. Compared to \optprefix{CUORE}0, the
larger array of CUORE affords more powerful time coincidence analysis
and more effective self-shielding from external backgrounds,
particularly those originating from the copper thermal shields or
cryostat. With this stronger background rejection and the already
demonstrated reduction of surface contamination, the CUORE background
goal of $0.01 \1{counts/(keV\cdot kg\cdot y)}$ is expected to be within reach.
The projected half-life sensitivity to $\n{^{130}Te}$ $0\nu\n{DBD}$ is
$9.5\times10^{25} \1{y}$ (90\% C.L.) with 5 years of live
time~\cite{Alessandria:2011rc}, reaching an effective Majorana
neutrino mass sensitivity of 0.05 to
$0.13\1{eV}$~\cite{Menendez:2008jp, PhysRevC.87.014301,
  Rodriguez:2010mn, Fang:2011da, Faessler:2012ku, suhonen_review_2012,
  Barea:2013bz, Kotila:2012zza}.

\section{Acknowledgments}
The CUORE Collaboration thanks the directors and staff of the
Laboratori Nazionali del Gran Sasso and the technical staff of our
laboratories. This work was supported by the Istituto Nazionale di
Fisica Nucleare (INFN); the Director, Office of Science, of the
U.S. Department of Energy under Contract Nos. DE-AC02-05CH11231 and
DE-AC52-07NA27344; the DOE Office of Nuclear Physics under Contract
Nos. DE-FG02-08ER41551 and DEFG03-00ER41138; the National Science
Foundation under Grant Nos. NSF-PHY-0605119, NSF-PHY-0500337,
NSF-PHY-0855314, NSF-PHY-0902171, and NSF-PHY-0969852; the Alfred
P. Sloan Foundation; the University of Wisconsin Foundation; and Yale
University. This research used resources of the National Energy
Research Scientific Computing Center (NERSC).

\bibliographystyle{apsrev} 
\bibliography{cuore0_performance}

\end{document}

%% file: author-20140202.tex
\author{D.~R.~Artusa\thanksref{USC,LNGS}
\and
F.~T.~Avignone~III\thanksref{USC}
\and
O.~Azzolini\thanksref{INFNLegnaro}
\and
M.~Balata\thanksref{LNGS}
\and
T.~I.~Banks\thanksref{BerkeleyPhys,LBNLNucSci,LNGS}
\and
G.~Bari\thanksref{INFNBologna}
\and
J.~Beeman\thanksref{LBNLMatSci}
\and
F.~Bellini\thanksref{Roma,INFNRoma}
\and
A.~Bersani\thanksref{INFNGenova}
\and
M.~Biassoni\thanksref{Milano,INFNMiB}
% ...
}

\thankstext{e1}{e-mail: cuore-spokeperson@lngs.infn.it}
\thankstext{d1}{Deceased}
\thankstext{g1}{Present address: Los Alamos National Laboratory, Los Alamos, NM, USA}

\institute{
Department of Physics and Astronomy, University of South Carolina, Columbia, SC 29208 - USA\label{USC}
\and
INFN - Laboratori Nazionali del Gran Sasso, Assergi (L'Aquila) I-67010 - Italy\label{LNGS}
\and
INFN - Laboratori Nazionali di Legnaro, Legnaro (Padova) I-35020 - Italy\label{INFNLegnaro}
\and
Department of Physics, University of California, Berkeley, CA 94720 - USA\label{BerkeleyPhys}
\and
Nuclear Science Division, Lawrence Berkeley National Laboratory, Berkeley, CA 94720 - USA\label{LBNLNucSci}
\and
INFN - Sezione di Bologna, Bologna I-40127 - Italy\label{INFNBologna}
\and
Materials Science Division, Lawrence Berkeley National Laboratory, Berkeley, CA 94720 - USA\label{LBNLMatSci}
\and
Dipartimento di Fisica, Sapienza Universit\`a di Roma, Roma I-00185 - Italy\label{Roma}
\and
INFN - Sezione di Roma, Roma I-00185 - Italy\label{INFNRoma}
\and
INFN - Sezione di Genova, Genova I-16146 - Italy\label{INFNGenova}
\and
Dipartimento di Fisica, Universit\`a di Milano-Bicocca, Milano I-20126 - Italy\label{Milano}
\and
INFN - Sezione di Milano Bicocca, Milano I-20126 - Italy\label{INFNMiB}
% ...
}